\documentclass[aps,prd,superscriptaddress,floatfix,showpacs]{revtex4}

\usepackage{verbatim}

\usepackage{amsmath}
\usepackage{amsfonts}
\usepackage{amssymb}
\usepackage{graphicx}
\usepackage{bm}
\usepackage{color}
\usepackage{epsf}
\usepackage{psfrag}
\newcommand{\beq}[1]{
\begin{equation}\label{#1}}
\newcommand{\eeq}{\end{equation}}
\newcommand{\bea}[1]{
\marginpar{\small\textsf{#1}}
\begin{eqnarray}\label{#1}}
\newcommand{\eea}{\end{eqnarray}}
\def\bea{\begin{eqnarray}}
\def\eea{\end{eqnarray}}
\def\beas{\begin{eqnarray*}}
\def\eeas{\end{eqnarray*}}
\def\beqas{\begin{eqnarray*}}
\def\eqas{\end{eqnarray*}}
\def\beq{\begin{equation}}
\def\eeq{\end{equation}}
\def\beqd{\begin{displaymath}}
\def\eeqd{\end{displaymath}}
\def\eqd{\end{displaymath}}

\def\slashchar#1{\setbox0=\hbox{$#1$}
   \dimen0=\wd0
   \setbox1=\hbox{/} \dimen1=\wd1
   \ifdim\dimen0>\dimen1
      \rlap{\hbox to \dimen0{\hfil/\hfil}}
      #1
   \else\begin{eqnarray}
      \rlap{\hbox to \dimen1{\hfil$#1$\hfil}}
      /
   \fi}

\begin{document}
\title
{Exclusive neutrino-production of a charmed vector meson and transversity gluon GPDs}
\author{ B.~Pire}
\affiliation{ Centre de Physique Th\'eorique, \'Ecole Polytechnique,
CNRS, 91128 Palaiseau,     France }

\author{ L.~Szymanowski}
\affiliation{ National Centre for Nuclear Research (NCBJ), Warsaw, Poland}

\date{\today}

\begin{abstract}

\noindent
We calculate at the leading order in $\alpha_s$ the QCD amplitude for exclusive  neutrino production of a  $D^*$ or  $D_s^*$ charmed vector meson on a nucleon. We work in the framework of the collinear QCD approach where generalized parton distributions (GPDs) factorize from  perturbatively calculable coefficient functions. We include  $O(m_c)$ terms in the coefficient functions and $O(m_D)$ term in the definition of  heavy meson distribution amplitudes. The show that the analysis of the angular distribution of the decay $D_{(s)}^*\to D_{(s)} \pi$  allows to access the transversity gluon GPDs.
\end{abstract}
%\pacs{}

\maketitle

\section{Introduction.}

The now well established framework of collinear QCD factorization \cite{fact1,fact2,fact3} for exclusive reactions mediated by a highly virtual photon in the generalized Bjorken regime describes hadronic amplitudes using generalized parton distributions (GPDs) which give access to a 3-dimensional analysis \cite{3d} of  the internal structure of hadrons.  
Neutrino production is another way to access (generalized) parton distributions  \cite{LDS,weakGPD,PS}. Although neutrino induced cross sections are orders of magnitudes smaller than those for electroproduction and neutrino beams are much more difficult to handle than charged lepton beams, they have  been very important to scrutinize the flavor content of the nucleon and the advent of new generations of neutrino experiments will open new possibilities. 

In \cite{PS}, we showed that exclusive neutrino production of a charmed scalar $D^+$ meson allows to access the transversity chiral-odd quark GPDs. In this paper we complement this study  and consider  the exclusive production of a vector $D^*(2010)$ or $D_s^*(2112)-$meson through the reactions:
\begin{eqnarray}
\nu_l (k)p(p_1, \lambda) &\to& l^- (k')D^{*+} (p_D,\varepsilon_D)p'(p_2, \lambda') \,,\\
\nu_l (k)n(p_1, \lambda) &\to& l^- (k')D^{*+} (p_D,\varepsilon_D)n'(p_2, \lambda') \,,
%\\
%\nu_l (k)n(p_1) &\to& l^- (k')D^{*0} (p_D,\varepsilon_D)p'(p_2) \,,\\
% \bar\nu_l (k) p(p_1) &\to& l^+ (k') D^{*-}(p_D,\varepsilon_D) p' (p_2)\,,\\
% \bar\nu_l (k) p(p_1) &\to& l^+ (k') D^{*0}(p_D,\varepsilon_D) n' (p_2)\,,\\
 % \bar\nu_l (k) n(p_1) &\to& l^+ (k') D^{*-}(p_D,\varepsilon_D) n' (p_2)\,,
\end{eqnarray}
in the kinematical domain where collinear factorization  leads to a description of the scattering amplitude 
in terms of 
%quark and 
gluonic \footnote{ we shall not consider here the quark contributions which are negligible in the $D_s^*$ case and not dominant at sufficiently large energy for $D^*$ production.} GPDs and the $D^*$ or $D_s^*-$meson distribution amplitude, with the hard subprocess  ($q=k-k', Q^2 = -q^2$):
\begin{eqnarray}
%W^+(q) d \to D^{*+} d~~~~&;&
~~~~ W^+(q) g \to  D^{*+} g\,,
\end{eqnarray}
described by the  handbag Feynman diagrams of Fig. 1.
% and Fig. 2..

%%%%%%%%%%%%%%%%%%%%%%%%%%%%%%%%%%%%%%%%%%%%%%%%%%%%%
\begin{figure}
\includegraphics[width=0.9\textwidth]{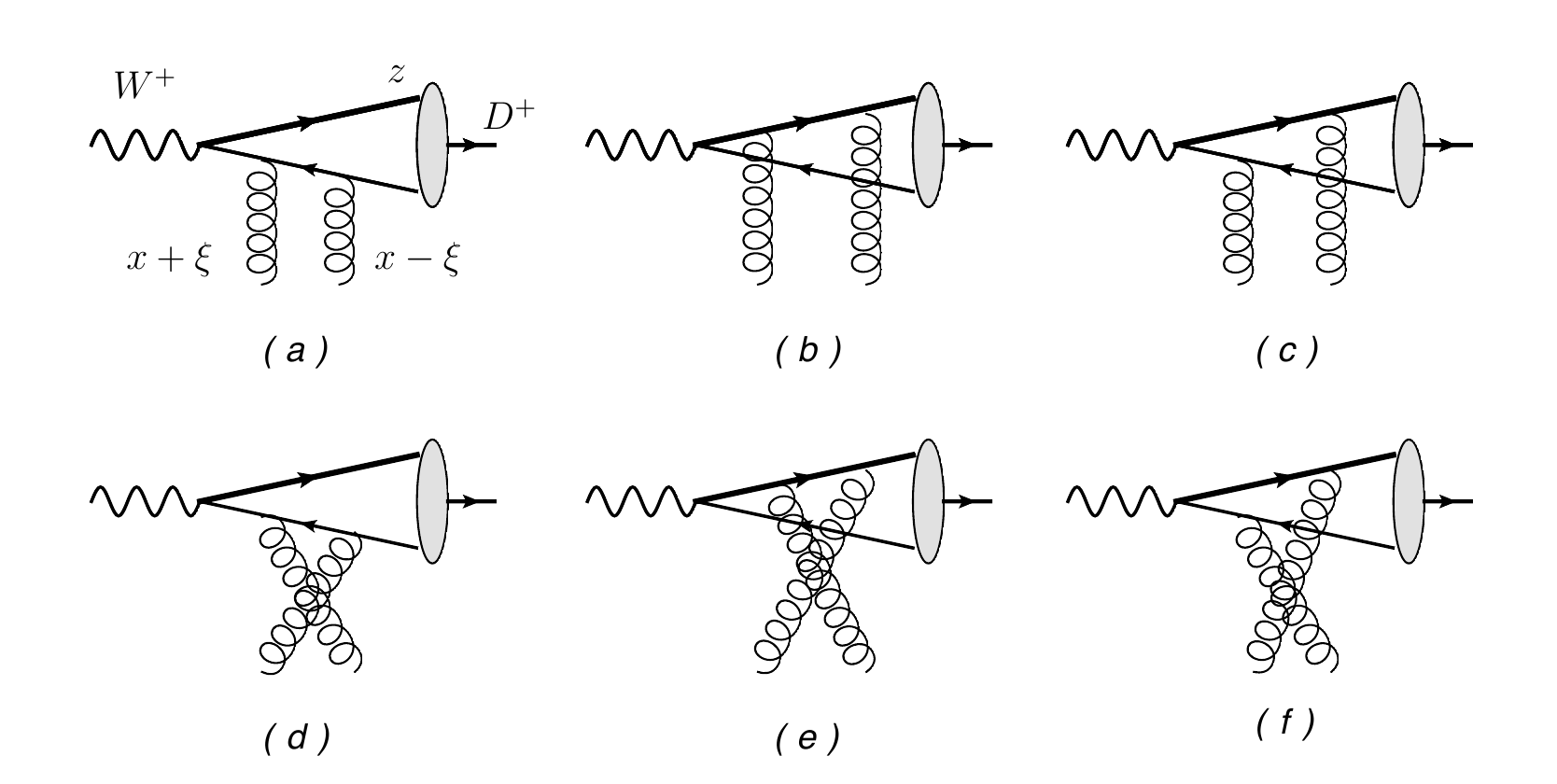}
%\vspace{1cm}
\caption{Feynman diagrams for the factorized  amplitude for the $ W^+ N \to  D^{*+} N'$ process involving the gluon GPDs; the thick line represents the heavy quark. The three other diagrams are obtained through the transformation ($x\to -x ; \mu \to \nu$).}
   \label{Fig1}
\end{figure}
Here we show that neutrino production of $D^*$ mesons may help to measure the gluon transversity GPDs, the phenomenology of which is presently restricted to angular asymmetries in DVCS  \cite{Belitsky:2000jk}, which turn out to be quite difficult to access experimentally.
%The longitudinal  amplitude $W_L ~q \to D^*(\epsilon_L) ~q$   gets its leading term in the collinear QCD framework as a convolution of chiral even leading twist quark GPDs and the corresponding gluon GPDs, with a  coefficient function of order  $O(\frac{1}{Q})$  while the transverse amplitude $W_T ~q \to D^*(\epsilon_T) ~q$   and $W_T ~g \to D^*(\epsilon_T) ~g$ gets its leading contribution of order $\frac{m_c}{Q^2}$ \cite{PS} from the convolution of transversity leading twist quark or gluon  GPDs with a coefficient function of order $m_c$. 
The charmed vector meson $D^*(2010)$ has been produced and studied in deep inelastic electroproduction at HERA \cite{D*atHERA}.
Its  main decay channel  is known to be
\begin{eqnarray}
D^*(2010) \to D(1870) \pi \,,
 \end{eqnarray}
and we will demonstrate that one may use the angular distribution of the final pion to disentangle various helicity amplitudes for $D^*$  production. The case of  $D_s^*(2112)$ is also interesting. It was already stressed by \cite{LDS} that the amplitude for $D_s(1968)-$neutrino-production is enhanced by the large value of the CKM matrix element $V_{sc}$ and is dominated by the contribution of gluon GPDs; its main decay channel  is:
\begin{eqnarray}
D_s^*(2112) \to D_s(1968) \gamma \,,
 \end{eqnarray}
but it turns out to be more difficult to use this decay channel  to disentangle various helicity amplitudes. 

\section{$D^*$ and $D_s^*$ meson distribution amplitudes} 
In the collinear factorization framework, the hadronization of the quark-antiquark pair is described by a distribution amplitude(DA) which obeys a twist expansion and evolution equations. Much work has been devoted to this subject \cite{heavyDA}.
The charmed meson distribution amplitudes are less known than the light meson ones. Here, we shall  restrict ourselves to a leading twist description of the $D^*$ meson DA but including the $M_D$ term in the definition of the correlator.   Omitting the path-ordered gauge link, the relevant distribution amplitudes read for the  $D^*$ vector meson case in the longitudinal or transverse polarization state:

\begin{eqnarray}
\langle D^{*+}(P_D, \varepsilon_L) | \bar c_\beta(y) d_\gamma(-y) |0 \rangle & =&
   \frac{f_L}{4} \int_0^1 dz e^{i(z-\bar z)P_D.y} [\hat P_D - M_D]_{\gamma \beta}   \phi_D(z)\,,
\\
\langle D^{*+}(P_D,\varepsilon_T) | \bar c_\beta(y) d_\gamma(-y) |0 \rangle & =&
   \frac{f_T}{4} \int_0^1 dz e^{i(z-\bar z)P_D.y} [ (\hat P_D - M_D)\hat \varepsilon_T]_{\gamma \beta}   \phi_D(z)\,,
         \end{eqnarray}
         where $\int_0^1 dz ~ \phi(z) = 1$.
 It has been argued that a heavy-light meson DA is strongly peaked around $z_0 = \frac {m_c}{M_D}$. We will parametrize  $\phi_D(z) $ in a very crude way, {\it i.e.} $\phi_D(z) = \delta(z-\frac{m_c}{M_D})$.

\section{Transversity gluon GPDs}

The four transversity gluon GPDs are defined in \cite{transGPDdef} as:

\begin{eqnarray}
  \label{defGPD}
&&- \frac{1}{P^+} \int \frac{d z^-}{2\pi}\, e^{ix P^+ z^-}
  \langle p',\lambda'|\, {\mathbf S}
     F^{+i}(-{\textstyle\frac{1}{2}}z)\,
     F^{+j}({\textstyle\frac{1}{2}}z)
  \,|p,\lambda \rangle \Big|_{z^+=0,\,\mathbf{z}_T=0} 
= {\mathbf S}\,
\frac{1}{2 P^+}\, \frac{P^+ \Delta^j - \Delta^+ P^j}{2 m P^+}
\nonumber \\
&\times& \bar{u}(p',\lambda') \left[
 H_T^g\, i \sigma^{+i} +
 \tilde H_T^g\, \frac{P^+ \Delta^i - \Delta^+ P^i}{m^2} 
%&& \left. \hspace{2.8em} {}+
 +E_T^g\, \frac{\gamma^+ \Delta^i - \Delta^+ \gamma^i}{2m} +
\tilde E_T^g\, \frac{\gamma^+ P^i - P^+ \gamma^i}{m}\, 
\right] u(p,\lambda) .
\end{eqnarray}
where $\mathbf S$ is an operator that symmetrizes a tensor and subtracts its trace : ${\mathbf S} t^{ij} = ( t^{ij} + t^{ji} ) /2  -  \delta^{ij} /2 *t^{kk}$. $\mathbf S$ may be written with the help of the $\tau^\perp_{\mu \nu; \rho \sigma}$ tensor defined as
\begin{equation}
\tau^\perp_{\mu \nu; \rho \sigma} = \frac{1}{2} [\,g^\perp_{\mu \rho} g^\perp_{\nu \sigma} +g^\perp_{\mu \sigma} g^\perp_{\nu \rho}-g^\perp_{\mu \nu} g^\perp_{ \rho\sigma}\,]\,,
\end{equation}
which appears in the Fierz decomposition of the Lorentz structure describing the two transverse gluon entering the coefficient function
\begin{equation}
g^\perp_{\mu \mu'} g^\perp_{\nu \nu'} =\frac{1}{2} g^\perp_{\mu \nu} g^\perp_{\mu' \nu'} + \frac{1}{2} \epsilon^\perp_{\mu \nu} \epsilon^\perp_{\mu' \nu'} + \tau^\perp_{\mu \nu; \rho \sigma} \tau^\perp_{\mu' \nu'; \rho \sigma}\,.
\end{equation}
As obvious from Eq. \ref{defGPD}, these transversity GPDs always appear accompanied by a factor $\Delta_T$ (at least) in an amplitude. To be consistent with the collinear factorization procedure, we calculate the coefficient function at zero $\Delta_T$. When computing a hadronic cross section, we restrict to lowest order in  $\Delta_T$, which means that we neglect all $\Delta_T$ effects in the definition of the polarization of the produced vector meson, and in the definition of its decay angles. 

There is no known parametrization for the four transversity gluon GPDs  \cite{models} and we shall not propose here a model for them. Since they do not have a forward non-zero limit, they cannot easily be constructed from a double distribution ansatz. Some inequalities have been derived \cite{Pob} based on positivity constraints \cite{PST}.

\section{The scattering amplitude}

\subsection{Kinematics}
Our kinematical notations are as follows ($m$ and $M_D$ are the nucleon and $D^*$meson masses):
\begin{eqnarray}
q=k-k'~~~~~; ~~~~~Q^2 = -q^2~~~~~&; &~~~~~P = \frac{p_1 +p_2}{2} ~~~~~; ~~~~~\Delta = p_2-p_1 ~~~~~; ~~~~~\Delta^2=t \,;\nonumber\\
p_1^\mu=(1+\xi)p^\mu +\frac{1}{2}  \frac{m^2}{1+\xi} n^\mu~~~~&;&~~~~ p_2^\mu=(1-\xi)p^\mu +\frac{1}{2}  \frac{m^2-\Delta_T^2}{1-\xi} n^\mu +\Delta_T^\mu\,; \\
q^\mu= -2\xi' p^\mu +\frac{Q^2}{4\xi'} n^\mu ~&;&~p_D^\mu=  2(\xi-\xi') p^\mu +\frac{M_D^2-\Delta_T^2}{4(\xi-\xi')}  n^\mu -\Delta_T^\mu \,,\nonumber
\end{eqnarray}
with $p^2 = n^2 = 0$ and $p.n=1$. Momentum conservation leads to the relation
\begin{equation}
\frac{Q^2}{\xi'}-\frac{2(m^2-\Delta_T^2)}{1-\xi} = \frac{M_D^2-\Delta_T^2}{\xi-\xi'} \,.
\end{equation}
As in  the double deeply virtual Compton scattering case~\cite{DDVCS}, it is meaningful to introduce two distinct momentum fractions:
 \begin{eqnarray}
\xi = - \frac{(p_2-p_1).n}{2} ~~~~~,~~~~~\xi' = - \frac{q.n}{2} \,.
\end{eqnarray} 
Neglecting the nucleon mass and $\Delta_T$, the approximate values of $\xi$ and $\xi'$ are
\begin{eqnarray}
\xi \approx \frac{Q^2+M_D^2}{4p_1.q-Q^2-M_D^2} ~~~~,~~~  \xi' \approx \frac{Q^2}{4p_1.q-Q^2-M_D^2} \,.
\end{eqnarray}
so that $\frac{Q^2}{\xi'}=\frac{Q^2+M_D^2}{\xi}$. We note $\kappa = p\cdot p_D =\frac{M_D^2}{4(\xi-\xi')}=\frac{M_D^2+Q^2}{4\xi}$.

\subsection{The  coefficient function}

The   coefficient function is calculated from the Feynman diagrams of Fig.\ref{Fig1} (the three  diagrams (d, e, f) are obtained from (a, b, c) by a ($x\to -x ; \mu \to \nu$) operation. Since we use a simple $\delta(z-m_c/M_D^*)$ form  for the $D^*$ distribution amplitude, we put $z = m_c/M_D^*$ to simplify    these coefficient functions.

\vspace{1 em}

\noindent
{\em Diagrams $a$ and $d$}.

The vector part of the diagrams of Fig.\ref{Fig1} (a and d)  contributes to the longitudinal amplitude and produce a longitudinally polarized $D^*$ meson. 
\begin{eqnarray}
G^{ad}_L =
%&=&-8 M (z-1) \epsilon \cdot \text{$\epsilon$W} \left(M^2 (z-1)-\kappa  (2 \xi +\text{xmxi})\right) / D_1(x,\xi)D_2(x,\xi) \nonumber \\
%   & = &  - 8M \bar{z}  ~\epsilon \cdot \epsilon_W \frac{M^2 \bar{z} +\kappa  (x+\xi)}{ D_1(x,\xi) D_2(x,\xi)} + \{ x \to -x  \}\,,
 \frac{- 8 M \bar z [ \epsilon_D \cdot
   \epsilon_W (\kappa 
   (x-3\xi )+M^2
   \bar z ) +2 \xi (x+ \xi )
   p\cdot \epsilon_D ~ p \cdot \epsilon_W ]} { D_1(x,\xi) D_2(x,\xi)} + ( x \to -x ) \,.
%8 M (z-1) \left(\epsilon \cdot \text{$\epsilon $W} \left(\kappa    (\text{xmxi}-2 \xi )-M^2   (z-1)\right)+(2 \xi +\text{xmxi})   p\cdot \epsilon  \text{pD}\cdot   \text{$\epsilon $W}\right)
     \end{eqnarray}
   It also contributes to the transverse amplitude  producing a transversally polarized $D^*$ meson:
\begin{eqnarray}
G^{ad}_T =
%&=&-8 M (z-1) \epsilon \cdot \text{$\epsilon$W} \left(M^2 (z-1)-\kappa  (2 \xi +\text{xmxi})\right) / D_1(x,\xi)D_2(x,\xi) \nonumber \\
%   & = &  - 8M \bar{z}  ~\epsilon \cdot \epsilon_W \frac{M^2 \bar{z} +\kappa  (x+\xi)}{ D_1(x,\xi) D_2(x,\xi)} + \{ x \to -x  \}\,,
 \frac{- 8 M \bar z \epsilon_D \cdot
   \epsilon_W \left(\kappa 
   (x-3\xi )+M^2
   \bar z\right) } { D_1(x,\xi) D_2(x,\xi)} + \{ x \to -x  \}\,,
  \end{eqnarray}
   and to the transverse scattering amplitude via the $\epsilon^{ijpn}$ part of the coefficient function  which is convoluted with $\tilde H$ and $\tilde E$ gluon GPDs:
 \begin{eqnarray}
\tilde G_{T}^{ad}&=& 
%\frac{- 8 M \bar z \epsilon ^{np\epsilon   \epsilon_W} \left(M^2   \bar z +\kappa  (x+ \xi   )\right)} { D_1(x,\xi) D_2(x,\xi)} + \{ x \to -x  \}\,,\\
%   &&  -8 M (z-1) \epsilon ^{np\epsilon   \text{$\epsilon $W}} \left(M^2   (z-1)+\kappa  (2 \xi   -\text{xmxi})\right) \\
   \frac{ -8 M \bar z \epsilon ^{np\epsilon_D \epsilon_W} \left(M^2 \bar z +\kappa  (x-3 \xi)\right) } { D_1(x,\xi) D_2(x,\xi)} - \{ x \to -x  \}\,, \, .
   \end{eqnarray}

The axial part \footnote{The traces which include a $\gamma^5$ matrix are straightforwardly calculated by using the fact that $P_D$ and $p$ are orthogonal to all other vectors so that 
$\gamma^5\, \hat P_D\, \hat p\, \hat \varepsilon_D  = P_D.p \,\gamma^5  \hat  \varepsilon_D -i \epsilon^{ P_D  p \,\varepsilon_D \rho} \gamma_\rho$
and
$ Tr [ \gamma^5\, \hat P_D\, \hat p\, \hat \varepsilon_D \gamma^\mu \gamma^\nu \hat \varepsilon] \tau^\perp_{\mu \nu; \rho \sigma}= -i \epsilon^{ P_D  p \,\varepsilon_D \rho} Tr [\gamma_\rho \gamma^\mu \gamma^\nu \hat \varepsilon] \tau^\perp_{\mu \nu; \rho \sigma}= -4i g_\perp^{\mu \nu} \epsilon^{ P_D  p \,\varepsilon_D \varepsilon}\tau^\perp_{\mu \nu; \rho \sigma}\, .
$} of the diagrams of Fig.\ref{Fig1} (a and d)  contributes with the $H(x,\xi)$ and $E(x,\xi)$  GPDs to the transverse amplitude and gives
\begin{eqnarray}
G_{5T}^{ad}
   &=&
   \frac{ 8 i M \bar z \kappa (x+\xi))
   \epsilon ^{p n \epsilon_D
\epsilon_W }}{D_1(x,\xi) D_2 (x,\xi)} + \{ x \to -x  \} \,.
\end{eqnarray}

It also contributes with the $\tilde H(x,\xi)$ and $\tilde E(x,\xi)$  GPDs to the longitudinal amplitude  
 \begin{eqnarray}
\tilde G_{5L}^{ad} = 
%8 i M (z-1) (2 \xi +\text{xmxi}) p\cdot \text{pD} \epsilon \cdot   \text{$\epsilon $W}
% -  \frac{8i \bar z M (x+\xi) ~p \cdot p_D  ~ \epsilon \cdot  \epsilon_W }{D_1(x,\xi) D_2 (x,\xi)} - \{ x \to -x  \}\, .
\frac{ 8 i M \bar z~ p\cdot \epsilon_W ~p\cdot \epsilon_D  \left(M^2 (\xi-x-2\xi z)+4 \kappa~ \xi ~(x-  \xi) \right)  }{\kappa~ D_1(x,\xi) D_2 (x,\xi)} - \{ x \to -x  \} \,,
  \end{eqnarray}
and to the transverse amplitude
 \begin{eqnarray}
\tilde G_{5T}^{ad} = 
%8 i M (z-1) (2 \xi +\text{xmxi}) p\cdot \text{pD} \epsilon \cdot   \text{$\epsilon $W}
% -  \frac{8i \bar z M (x+\xi) ~p \cdot p_D  ~ \epsilon \cdot  \epsilon_W }{D_1(x,\xi) D_2 (x,\xi)} - \{ x \to -x  \}\, .
  \frac{ 8 i M_D \bar z \kappa (x+\xi )    \epsilon_D \cdot \epsilon_W }{D_1(x,\xi) D_2 (x,\xi)} - \{ x \to -x  \} \,.
  \end{eqnarray}

\vspace{1 em}

\noindent
{\em Diagrams $b$ and $e$}.

The vector part of diagram  \ref{Fig1} (b and e) gives a  contribution to  the longitudinal amplitude :
\begin{eqnarray}
G^{be}_L
%&=& -16 \kappa  z M \text{xmxi}  \epsilon \cdot  \text{$\epsilon $W} / D_3(x,\xi) D_4(x,\xi) \nonumber \\
   &=&-16 \frac {\kappa  z M (x-\xi)  \epsilon \cdot \epsilon_W }{D_3(x,\xi) D_4(x,\xi)} + \{ x \to -x  \}\,,
\end{eqnarray}
and to the transverse amplitude:
\begin{eqnarray}
G^{be}_T
%&=& -16 \kappa z M \text{xmxi}  \epsilon \cdot   \text{$\epsilon $W} \\
   &=&-16 \frac { \kappa z M (x-\xi) \epsilon \cdot \epsilon_W }{D_3(x,\xi) D_4(x,\xi)} + \{ x \to -x  \}\,,
\end{eqnarray}
but not  to the transverse scattering amplitude via the $\epsilon^{ijpn}$ part of the coefficient function  which is convoluted with $\tilde H$ and $\tilde E$ gluon GPDs.
 \begin{eqnarray}
\tilde G_{T}^{be}&=& \frac{16 \kappa  M z (\xi -x) \epsilon
   ^{np\epsilon \epsilon_W} }{D_3(x,\xi) D_4(x,\xi)} - \{ x \to -x  \} = 0 \,.
\end{eqnarray}
The axial part of diagram  \ref{Fig1} (b and e) gives a  zero contribution to the amplitude.
%via the $\epsilon^{ijpn}$ part of the coefficient function  which is convoluted with $\tilde H$ and $\tilde E$ gluon GPDs.
%\begin{eqnarray}
%\tilde G^{be}_{5L} &=&\frac{16 i M (x-\xi) z p_D \cdot \epsilon_W    p\cdot \epsilon}{D_3(x,\xi) D_4(x,\xi)} - \{ x \to -x  \} = 0\,.
%\end{eqnarray}

\vspace{1 em}

\noindent
{\em Diagrams $c$ and $f$}. 

The vector part of diagram  \ref{Fig1} (c and f) gives a  contribution to   the longitudinal amplitude :
\begin{eqnarray}
G^{cf}_L
%&=& -\frac{4 M \text{xmxi} p\cdot   \text{$\epsilon $W} \left(p\cdot   \epsilon  \left(M^2 (z-1)+4 \kappa    \text{xpxi}\right)-2 \kappa ^2 (z-1)   n\cdot \epsilon \right)}{\kappa} \nonumber \\
 &=& -\frac{4 M (x-\xi) p\cdot \epsilon_W \left(p\cdot
   \epsilon_D  \left(- M^2 \bar z +4 \kappa 
  (x+\xi)\right)+2 \kappa ^2 \bar z
   n\cdot \epsilon_D \right)}{\kappa D_4(x,\xi) D_2(-x,\xi)} + \{ x \to -x  \}\,,
\end{eqnarray}
and to the TT amplitude (to be convoluted with transversity GPDs)
\begin{eqnarray}
G^{cf}_{\tau}
%&=&-4 \kappa  M \text{xmxi} (z-1)   \left(\text{$\epsilon $W}^{\alpha }   \epsilon ^{\beta }+\epsilon ^{\alpha }   \text{$\epsilon $W}^{\beta }-g^{\alpha   \beta } \epsilon \cdot \text{$\epsilon   $W}\right)/ D_3(x,\xi) D_5(x,\xi) \nonumber \\
   &=& 4 \kappa  M (x-\xi) \bar z \frac{
   \epsilon_W^{\alpha}
   \epsilon ^{\beta }+\epsilon ^{\alpha }
  \epsilon_W^{\beta } - g_T^{\alpha
   \beta } \epsilon \cdot \epsilon_W}{D_4(x,\xi) D_2(-x,\xi)} + \{ x \to -x  \}\,,
\end{eqnarray}
 but not to the transverse amplitude
 \begin{eqnarray}
G^{cf}_{T}&=0 \,.
\end{eqnarray}
Moreover, it does not give any contribution to the $\epsilon^{ijpn}$ part of the amplitude which is convoluted with $\tilde H$ and $\tilde E$ gluon GPDs.
 \begin{eqnarray}
\tilde G^{cf}_{T}&=&0 \,.
\end{eqnarray}
The axial part of  diagram  \ref{Fig1} (c and f) convoluted to the $\tilde H$ and the $\tilde E$ GPDs gives a contribution to the longitudinal amplitude 
\begin{eqnarray}
&\tilde G^{cf}_{5L}&
%=-4 i M \text{xmxi} (p\cdot \text{$\epsilon   $D} (2 p\cdot \text{$\epsilon $W}   ((z-1) n\cdot \text{pD}+2   \text{xpxi})+3 (z-1) \text{pD}\cdot   \text{$\epsilon $W})-3 \kappa  (z-1)   n\cdot \text{$\epsilon $W} p\cdot  \text{$\epsilon $D}+\kappa  (z-1)   n\cdot \text{$\epsilon $D} p\cdot   \text{$\epsilon $W})\nonumber \\
%= -\frac{8 i M \text{xmxi} p\cdot   \text{$\epsilon $D} p\cdot   \text{$\epsilon $W}   (\text{M2} (z-1)+2 \kappa    (x+\xi ))}{\kappa } \\
   = -\frac{8 i M (x-\xi)~p\cdot
   \epsilon_D ~p\cdot
  \epsilon_W
   (- M^2 \bar z+2 \kappa 
   (x+\xi ))}{\kappa D_4(x,\xi) D_2(-x,\xi)}   - \{ x \to -x  \}\,.
 %  \\&=& \frac{-4 i M (x-\xi) \left[ p\cdot \epsilon_D (2 p\cdot \epsilon_W   (-\bar z n\cdot p_D+2   (x+\xi))-3 \bar z p_D\cdot   \epsilon_W)+3 \kappa  \bar z  n\cdot \epsilon_W  p\cdot  \epsilon_D-\kappa  \bar z   n\cdot \epsilon_D p\cdot   \epsilon_W\right]}{D_4(x,\xi) D_2(-x,\xi)}\nonumber \\
%&& - \{ x \to -x  \}\,.
\end{eqnarray}
It does not give any  contribution to the transverse amplitude 
but a  contribution to  the transversity GPD part of the TT amplitude 
%\begin{eqnarray}G_{5 \tau}^{cf}
%&=&2 i (M (\text{xmxi} (z-1)-\text{xpxi} z)+M   \text{xpxi} z) \left(g^{\alpha \beta }   p\cdot \text{$\epsilon $D} \epsilon   ^{np\text{pD}\text{$\epsilon   $W}}+g^{\alpha \beta } p\cdot   \text{$\epsilon $W} \epsilon   ^{np\text{pD}\text{$\epsilon $D}}+2   \text{$\epsilon $D}^{\alpha } \epsilon   ^{\beta p\text{pD}\text{$\epsilon   $W}}+2 \text{$\epsilon $W}^{\beta }   \epsilon ^{\alpha   p\text{pD}\text{$\epsilon $D}}+p\cdot   \text{$\epsilon $D} \epsilon ^{\alpha   \beta \text{pD}\text{$\epsilon   $W}}-p\cdot \text{$\epsilon $W}   \epsilon ^{\alpha \beta   \text{pD}\text{$\epsilon   $D}}+\text{pD}\cdot \text{$\epsilon $W}   \epsilon ^{\alpha \beta   p\text{$\epsilon $D}}\right)  / D_4(x,\xi) D_2(-x,\xi) \nonumber \\
 %  &=& \frac{- 2 i M \bar z (x-\xi)) \left(2  \epsilon_D^{\alpha } \epsilon   ^{\beta pp_D\epsilon_W}+2 \epsilon_W^{\beta }   \epsilon ^{\alpha   pp_D \epsilon_D}+p\cdot   \epsilon_D \epsilon ^{\alpha   \beta p_D\epsilon_W}-p\cdot \epsilon_W   \epsilon ^{\alpha \beta  p_D\epsilon_D}+p_D\cdot \epsilon_W   \epsilon ^{\alpha \beta   p \epsilon_D}\right) }{ D_4(x,\xi) D_2(-x,\xi)} \nonumber\\   && + \{ x \to -x  \}\, .   \end{eqnarray}
which reduces after using the fact that only the symmetric part in$(\alpha, \beta)$ contributes, to 
 \begin{eqnarray}
G_{5 \tau}^{cf}
%&=&2 i (M (\text{xmxi} (z-1)-\text{xpxi} z)+M   \text{xpxi} z) \left(g^{\alpha \beta }   p\cdot \text{$\epsilon $D} \epsilon   ^{np\text{pD}\text{$\epsilon   $W}}+g^{\alpha \beta } p\cdot   \text{$\epsilon $W} \epsilon   ^{np\text{pD}\text{$\epsilon $D}}+2   \text{$\epsilon $D}^{\alpha } \epsilon   ^{\beta p\text{pD}\text{$\epsilon   $W}}+2 \text{$\epsilon $W}^{\beta }   \epsilon ^{\alpha   p\text{pD}\text{$\epsilon $D}}+p\cdot   \text{$\epsilon $D} \epsilon ^{\alpha   \beta \text{pD}\text{$\epsilon   $W}}-p\cdot \text{$\epsilon $W}   \epsilon ^{\alpha \beta   \text{pD}\text{$\epsilon   $D}}+\text{pD}\cdot \text{$\epsilon $W}   \epsilon ^{\alpha \beta   p\text{$\epsilon $D}}\right)  / D_4(x,\xi) D_2(-x,\xi) \nonumber \\
   &=& \frac{- 4 i M \kappa \bar z (x-\xi) \left(
   \epsilon_D^{\alpha } \epsilon
   ^{\beta pn \epsilon_W} + \epsilon_W^{\beta }
   \epsilon ^{\alpha
   pn \epsilon_D}\right) }{ D_4(x,\xi) D_2(-x,\xi)}  + \{ x \to -x  \}\, .
   \end{eqnarray}

The denominator of the quark propagators are given by :
\begin{eqnarray}
D_1 (x,\xi)&=&- \bar z[z M_D^2 + Q^2]+i\epsilon\,, \nonumber \\
D_2(x,\xi)&=&  \bar z (Q^2+M_D^2) \frac{ x- \xi +\bar z \alpha \xi }{2\xi} +i\epsilon = 2  \bar z \tau ( x- \xi +\bar z \alpha \xi)+i\epsilon \,, \nonumber \\
D_3 (x,\xi)&=&  - z [M_D^2 + Q^2] +i\epsilon = - 4 \xi \tau z  +i\epsilon\,,  \\
D_4 (x,\xi)&=&  z [M_D^2 + Q^2] \frac{x-\xi}{2 \xi}+i\epsilon = 2  z \tau ( x- \xi)+i\epsilon\,, \nonumber 
 %D_5 &=& \bar z^2 M_D^2 - \bar z (Q^2+M_D^2) \frac{x+\xi-i\epsilon}{2\xi} = -\bar z (Q^2+M_D^2) \frac{x+\xi -\bar z \alpha \xi }{2\xi} +i\epsilon\,, 
 %\nonumber \\
 %D'_1 &=& (Q^2+M_D^2) (z\frac{x-\xi}{2 \xi}+  \frac{z^2 M_D^2-m_c^2}{Q^2+M_D^2} +i\epsilon) = z (Q^2+M_D^2)\frac{x-\xi +\beta \xi + i\epsilon}{2 \xi} \,, \nonumber \\
 %D'_2 &=& \bar z^2 M_D^2 - \bar z (Q^2+M_D^2) \frac{x+\xi-i\epsilon}{2\xi} = -\bar z (Q^2+M_D^2) \frac{x+\xi -\bar z \alpha \xi -i\epsilon}{2\xi}\,, \nonumber \\
   %D'_3 &=& (Q^2+M_D^2) (-z+  \frac{z^2 M_D^2-m_c^2}{Q^2+M_D^2} ) \approx - z (Q^2+M_D^2)\, ,
 \label{Denom}
\end{eqnarray}
 where  $\alpha = \frac {2 M_D^2}{Q^2+M_D^2}$.
 % and $\beta = \frac {2 z (M_D^2-m_c^2/z^2)}{Q^2+M_D^2}\approx 0$.

These computations show that there are five non-zero ($W \to D^*$) helicity amplitudes:
\begin{itemize}
\item a longitudinal ($W$) to longitudinal ($D^*$) amplitude ${\cal M}_{00}$, originated from $G_L= G_L^{ad}+ G_L^{be} +  G_L^{cf}$ and from $\tilde G_L= \tilde G_{5L}^{ad}+ \tilde G_{5L}^{cf} $;
\item a left ($W$) to left ($D^*$)  and a  right ($W$) to  right (($D^*$) amplitude, ${\cal M}_{LL}$ and  ${\cal M}_{RR}$, originated from $G_T= G_T^{ad}+ G_{5T}^{ad} +  G_T^{be}$ and from $\tilde G_T= \tilde G_T^{ad}+ \tilde G_{5T}^{ad} $;
\item a left ($W$) to right ($D^*$) and a  right ($W$) to left ($D^*$) amplitude, ${\cal M}_{LR}$ and ${\cal M}_{RL}$, originated from $G_\tau^{\alpha \beta}= G_\tau^{cf}+   G_{5\tau}^{cf}$, which are proportional to transversity gluon GPDs.
\end{itemize}

 These  amplitudes read:
\begin{eqnarray}
 {\cal M}_{00}&=&  \frac{ i C_g}{2} \int_{-1}^{1}dx \frac{- 1}{(x+\xi-i\epsilon)(x-\xi+i\epsilon)} \int_0^1 dz f_L \phi_{D^*}(z) \cdot \nonumber \\
&& \left[  \bar{N}(p_{2})[H\hat n +E\frac{i\sigma^{n\Delta}}{2m} ]N(p_{1}) G_L+\bar{N}(p_{2})[{\tilde H} \hat n \gamma^5+{\tilde E}\frac{\gamma^5n.\Delta}{2m} ]N(p_{1}) \tilde G_L  \right] \, ,
\end{eqnarray}
\begin{eqnarray}
 {\cal M}_{RR}={\cal M}_{LL}&=&  \frac{ i C_g}{2} \int_{-1}^{1}dx \frac{- 1}{(x+\xi-i\epsilon)(x-\xi+i\epsilon)} \int_0^1 dz f_T \phi_{D^*}(z) \cdot \nonumber \\
&& \left[  \bar{N}(p_{2})[H\hat n +E\frac{i\sigma^{n\Delta}}{2m} ]N(p_{1}) G_T+\bar{N}(p_{2})[{\tilde H} \hat n \gamma^5+{\tilde E}\frac{\gamma^5n.\Delta}{2m} ]N(p_{1}) \tilde G_T  \right] \, ,
\end{eqnarray}
%\begin{eqnarray}
% {\cal M}_{RR}&=&  \,, 
%\end{eqnarray}
\begin{eqnarray}
 {\cal M}_{LR}&=&   i C_g \int_{-1}^{1}dx \frac{- 1}{(x+\xi-i\epsilon)(x-\xi+i\epsilon)} \int_0^1 dz f_T \phi_{D^*}(z) G_\tau^{\alpha \beta} ~ {\mathbf S} \frac{P \cdot n \Delta^\beta -\Delta\cdot n P^\beta}{2m}\cdot \nonumber \\
&& \left[  \bar{N}(p_{2})[H_T i \sigma^{n\alpha}  +E_T\frac{\hat n \Delta^\alpha-\Delta \cdot n \gamma^\alpha}{2m} + \tilde H_T \frac{P \cdot n \Delta^\alpha - \Delta \cdot n P^\alpha}{m^2}  - \tilde E_T  \frac{P \cdot n \gamma^\alpha - \hat n P^\alpha}{m}]N(p_{1})    \right] \, 
\end{eqnarray}
\begin{eqnarray}
 {\cal M}_{RL}&=&  {\cal M}_{LR}^* \,, 
\end{eqnarray}
 with $T_f=\frac{1}{2}$ and the factor $\frac{- 1}{(x+\xi-i\epsilon)(x-\xi+i\epsilon)}$ comes from the conversion of the strength tensor to the  gluon field. $C_g= T_f\frac{\pi}{3} \alpha_s V_{dc}$ for $D^*$ production and $C_g= T_f\frac{\pi}{3} \alpha_s V_{sc}$ for $D_s^*$ production.

 \section{Observables.}  
   The differential cross section for neutrino production of a  charmed meson is written as \cite{Arens} :

\begin{eqnarray}
\frac{d^4\sigma(\nu_l N\to l^- N'D^*)}{dy\, dQ^2\, dt\,  d\varphi} &=&
\bar \Gamma\Biggl\{\frac{1}{2}(\sigma_{RR}^{(X)}+\sigma_{LL}^{(X)})+\varepsilon
\sigma_{00}^{(X)} -\varepsilon\cos(2\varphi)\ {\cal R}e\sigma_{RL}^{(X)}+\varepsilon\sin(2\varphi)\ {\cal I}m\sigma_{RL}^{(X)}\nonumber\\
&&-\sqrt{\varepsilon(1+\varepsilon)}\cos\varphi\
{\cal R}e(\sigma_{R0}^{(X)}-\sigma_{L0}^{(X)})+\sqrt{\varepsilon(1+\varepsilon)}\sin\varphi\
 {\cal I}m(\sigma_{R0}^{(X)}
+\sigma_{L0}^{(X)})\nonumber\\
&&-\sqrt{1-\varepsilon^2}\frac{1}{2}
(\sigma_{RR}^{(X)}-\sigma_{LL}^{(X)})
+\sqrt{\varepsilon(1-\varepsilon)}\cos\varphi\ {\cal R}e
(\sigma_{R0}^{(X)}+\sigma_{L0}^{(X)})
\nonumber\\
&&-\sqrt{\varepsilon(1-\varepsilon)}\sin\varphi\ {\cal I}m
(\sigma_{R0}^{(X)}-\sigma_{L0}^{(X)})\Bigr\}  \, .
\end{eqnarray}
 Here  $\sigma_{RR}= \sigma_{LL}$.  When integrated over the leptonic azimuthal angle $\varphi$, this yields   
 \begin{eqnarray}
\label{cs}
\frac{d^4\sigma(\nu N\to l^- N'D^*)}{dy\, dQ^2\, dt}
%&=&
%-\varepsilon\cos(2\varphi)\sigma_{+-}
%\\
 = 2 \pi \bar\Gamma
\Bigl\{ ~\sigma_{LL} &+&\varepsilon\sigma_{00} \Bigr\},
\end{eqnarray}
with $y= \frac{p \cdot q}{p\cdot k}$ , $Q^2 = x_B y (s-m^2)$, $\varepsilon \approx \frac{1-y}{1-y+y^2/2}$ and
\begin{equation}
\bar \Gamma = \frac{G_F^2}{(2 \pi)^4} \frac{1}{16y} \frac{1}{\sqrt{ 1+4x_B^2m_N^2/Q^2}}\frac{1}{(s-m_N^2)^2} \frac{Q^2}{1-\epsilon}\,, \nonumber
\end{equation}
where the ``cross-sections'' $\sigma_{lm}=\epsilon^{* \mu}_l W_{\mu \nu} \epsilon^\nu_m$ are product of  amplitudes for the process $ W(\epsilon_l) N\to D^* N' $, averaged  (summed) over the initial (final) nucleon polarizations.

The   $D^*$ meson  decays preferably through the $D \pi$ channel. The azimuthal distribution of this $D \pi$ final state (in the $D^*$ rest system) is obtained from the helicity matrix elements of the $D^*$  \cite{ Leader, Diehl:2007jy} related to the amplitudes for the production of a $D^*$ with definite helicity. 
The cross section integrated over the leptonic azimuthal angle but not over the ($\theta_D, \varphi_D$) decay angles thus reads ($B^{D^*\to X}$ is the branching ratio for the particular decay mode)
 \begin{eqnarray}
\label{csdec}
\frac{d^5\sigma(\nu N\to l^- N'D \pi)}{dy\, dQ^2\, dt d\theta_D d\varphi_D}
%&=&
%-\varepsilon\cos(2\varphi)\sigma_{+-}
%\\
 = 2 \pi \bar\Gamma B^{D^*\to D \pi} \frac{3}{8\pi}&
\Bigl\{&\varepsilon\sigma_{00} 2 \cos^2 \theta_D +  \sigma_{LL} \sin^2 \theta_D (1-\alpha \cos 2\varphi_D + \beta \sin 2\varphi_D) \Bigr\}\, ,
\end{eqnarray}
with
 \begin{eqnarray}
\label{albet}
\alpha &=& \frac{2~{\cal R}e [{\cal M}_{RR} {\cal M}_{RL}^*] }{\sigma_{LL}}\,~~~,~~~
\beta = \frac{2~{\cal I}m [{\cal M}_{RR} {\cal M}_{RL}^*] }{\sigma_{LL}}\,.
\end{eqnarray}

The fact that ${\cal M}_{LR}$ and thus $\alpha, \beta$ are linear in the transversity GPDs allows a direct access to them through the moments
  \begin{eqnarray}
\label{moments}
<\cos 2\varphi_D>&=& \frac { \int d\varphi_D  \cos 2\varphi_D \frac{d^5\sigma(\nu N\to l^- N'D \pi)}{dy\, dQ^2\, dt d\theta_D d\varphi_D}}{ \int d\varphi_D   \frac{d^5\sigma(\nu N\to l^- N'D \pi)}{dy\, dQ^2\, dt d\theta_D d\varphi_D}} \,,\\
<\sin 2\varphi_D>&=& \frac { \int d\varphi_D  \sin 2\varphi_D \frac{d^5\sigma(\nu N\to l^- N'D \pi)}{dy\, dQ^2\, dt d\theta_D d\varphi_D}}{ \int d\varphi_D   \frac{d^5\sigma(\nu N\to l^- N'D \pi)}{dy\, dQ^2\, dt d\theta_D d\varphi_D}} \,.
\end{eqnarray}

The   $D_s^*$ meson  decays preferably through the $D_s \gamma$ channel. The azimuthal distribution of such a pseudoscalar meson + photon state is model-dependent \cite{  Titov:2008hp}, and we shall not discuss it further. The suppressed decay $D^*_s \to D_s \pi$ may be treated in the same way as above.

\section{Conclusion.}
We have derived a new way to get access to the transversity gluon GPDs, the knowledge of which would shed a new light on the deep structure of the nucleons. $D^*$ mesons are produced in a medium energy neutrino experiment, at least in their longitudinal modes, at a rate comparable to the pseudoscalar $D$ mesons \cite{PS}. We however believe that it is too premature to work on the phenomenology of the proposed reaction. In the one hand, neutrino experiments did not yet demonstrate their ability to analyze exclusive production of a charmed meson, even a simple pseudoscalar $D$ meson. Analyzing the decay products of the $D^*$ meson is likely to be a formidable task since neutrino experiments cannot pretend to as precise energy reconstruction as electroproduction experiments. In the second hand, the non existence  of a reasonable model for transversity gluon GPDs would nullify any effort to predict the size of the angular modulation present in Eq.\ref{csdec}.
  
 Although planned high energy neutrino facilities \cite{NOVA} which have their scientific program oriented toward the understanding of neutrino oscillations   should allow  some important progress in the realm of hadronic physics, we cannot claim that the experimental measurement of the observables proposed here will be  feasibe. 
%%%%%%%%

%%%%%%%%%%%%%%%%%%%%%%%%%%%%%%%%%%%%%%%%%%%%%%%%%%%%%%%%%%%%%%%%%%%%%%%%%%% 
\paragraph*{Acknowledgements.}
\noindent
 We acknowledge useful correspondence with Markus Diehl. This work is partly supported by Grant No. 2015/17/B/ST2/01838 of the National Science Center in Poland, by the Polish-French Polonium agreements and by the COPIN-IN2P3 Agreement..

\end{document}